\documentclass{article}

\usepackage{ifthen}
\newboolean{neurips_version}
\setboolean{neurips_version}{true}
\ifthenelse{\boolean{neurips_version}}
{\PassOptionsToPackage{numbers, compress}{natbib}
 \usepackage[preprint]{neurips_2022}
}
{
\usepackage{PRIMEarxiv}
 \usepackage[numbers]{natbib}
}

\usepackage[utf8]{inputenc} 
\usepackage[T1]{fontenc}    
\usepackage{hyperref}       
\usepackage{url}            
\usepackage{booktabs}       
\usepackage{amsfonts,amsmath,amsthm,amssymb}       
\usepackage{nicefrac}       
\usepackage{microtype}      
\usepackage{xcolor}         
\usepackage[disable]{todonotes}
\usepackage{enumitem}
\usepackage{caption}

\usepackage{mathtools} 
\usepackage{tikz}
\usetikzlibrary{calc}
\usepackage{pgfplots}
\usepackage{graphicx}

\usepackage{amsmath}
\usepackage{algorithm, algpseudocode}
\usepackage{amsfonts}  

\usepackage[english]{babel}
\usepackage{amsthm}

\usepackage{subfig}
\usepackage[export]{adjustbox}

\usepackage{hhline}
\usepackage{makecell, caption, booktabs}
\usepackage{siunitx}

\usepackage{multirow}

\usepackage{amsmath, nccmath}
\usepackage{bigstrut}

\usepackage{booktabs}

\usepackage{lipsum}
\graphicspath{{media/}}     

\usepackage{CJKutf8}
\usepackage[framed,numbered,autolinebreaks,useliterate]{mcode}

\title{ Privacy-Preserving 3-Layer Neural Network Training }

\author{ \href{https://orcid.org/0000-0003-0378-0607}{\includegraphics[scale=0.06]{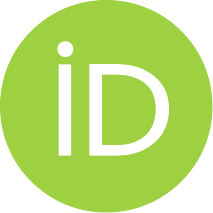}\hspace{1mm}John Chiang} \\                             
                                      \\
	\texttt{john.chiang.smith@gmail.com} 
}

\date{}

\theoremstyle{remark}

\renewcommand{\epsilon}{\varepsilon}

\makeatletter
\def\namedlabel#1#2{\begingroup
    #2%
    \def\@currentlabel{#2}%
    \phantomsection\label{#1}\endgroup
}
\makeatother

\algnewcommand{\LeftComment}[1]{\Statex \(\triangleright\) #1}
\algnewcommand{\LineCommentStep}[1]{\Statex \textbf{[Step #1]:} }
\makeatletter
\newlength{\trianglerightwidth}
\algnewcommand{\LineComment}[1]{\Statex \hskip\ALG@thistlm $\triangleright$ #1}
\algnewcommand{\LineCommentCont}[1]{\Statex \hskip\ALG@thistlm%
  \parbox[t]{\dimexpr\linewidth-\ALG@thistlm}{\hangindent=\trianglerightwidth \hangafter=1 \strut$\triangleright$ #1\strut}}
\algnewcommand{\LeftLineCommentCont}[1]{\Statex \hskip\ALG@thistlm%
  \parbox[t]{\dimexpr\linewidth-\ALG@thistlm}{\leftskip=\algorithmicindent \hangindent=\trianglerightwidth \hangafter=1 \strut$\triangleright$ #1\strut}}

\newenvironment{mbmatrix}{\begin{medsize}\begin{bmatrix}}{\end{bmatrix}\end{medsize}}

\newcommand{\mysplit}[1]{%
  \begin{tabular}{@{}c@{}}
    #1
  \end{tabular}
  }
  
\begin{document}

\maketitle

\begin{abstract}%
In this manuscript, we consider the problem of privacy-preserving training of neural networks in the mere homomorphic encryption setting. We combine several exsiting techniques available, extend some of them, and finally enable the training of 3-layer neural networks for both the regression and classification problems using mere homomorphic encryption technique.

\todo{Show precise somewhere that our results are new in the iid setting as well.}

\end{abstract}

\listoftodos

\section{Introduction}
\subsection{Background}
In the age of artificial intelligence and machine learning, neural networks have emerged as state-of-the-art models, exhibiting exceptional predictive capabilities for both regression and classification tasks. These networks are favored across diverse domains like healthcare and finance due to their remarkable performance. However, achieving high accuracy in training neural network models demands access to substantial volumes of private and sensitive data. This creates a requirement for individuals and institutions to securely share sensitive data, thereby extracting valuable insights from it.
Homomorphic Encryption emerges as a straightforward solution for preserving privacy during neural network training in such scenarios, offering a high level of security.

\subsection{Related Work}
Several papers already discussed privacy-preserving training of binary logistic regression~\citep{IDASH2018Andrey, IDASH2018bonte, IDASH2018chen, IDASH2018gentry, IDASH2019blatt, IDASH2019kim, chiang2022privacy, han2018efficient} and homomorphic inference on neural networks~\citep{chiang2022novel, chou2018faster, kim2018matrix}.  Sav et al.~\citep{sav2020poseidon} address the challenge of preserving privacy during the training and evaluation of neural networks within a multi-party federated learning framework.

The research most closely related to this paper is the work of Chiang et al.~\citep{chiang2023privacy, nandakumar2019towards}.   They also employed machine learning based on homomorphic encryption but used a faster gradient variant called Quadratic Gradient~\citep{chiang2022privacy, chiang2022quadratic, chiang2023multinomial} extending from the (Simplified) Fixed Hessian~\citep{IDASH2018bonte, bohning1988monotonicity, bohning1999lower, bohning1992multinomial}. However, it's important to note that in their work they only completed the  training for classification problems using multiclass logisitic regression, namely a simple 2-layer neural network without any hidden layer. 

\subsection{Contributions}
In this study, we extend homomorphic multiclass logistic regression training in the work~\citep{chiang2023privacy} to 3-layer neural networks with one single hidden layer of various numbles of nodes, for both regression and classification problems. We demonstrate the feasibility of training a homomorphic 3-layer neural network in the HE-encrypted domain and extend the loss function, Squared Likelihood Error, to its two variants that can be used for classification problems in privacy-preserving neural network training using mere homomorphic encryption. The Mean Squared Error (MSE) loss function is HE-friendly and can be applied for regression tasks in this paper.



\section{Preliminaries}
Let ``$\odot$''  denote the component-wise multiplication between matrices. The Sigmoid function is represented as follows: $$ Sigmoid (x) = \frac{  1 }{ 1 + \exp (- x ) } . $$.
 

\subsection{Fully Homomorphic Encryption}
Homomorphic encryption is an important technique in cryptography and is hailed as the Holy Grail in the field of cryptography. It is a special encryption method that allows calculations to be performed in an encrypted state without the need to decrypt the data. This means that by using homomorphic encryption, it is possible to perform various operations on encrypted data, such as addition, multiplication, etc., without having to decrypt them first. This provides a powerful tool for protecting data privacy, as the data remains encrypted at all times. 

HE is especially useful for handling sensitive data, such as medical records and financial information. However, due to the computational complexity of homomorphic encryption, it can be slower than traditional encryption methods. In recent years, with the continuous development of technology~\cite{gentry2009fully,brakerski2014leveled,SmartandVercauteren_SIMD,cheon2017homomorphic}, researchers have continuously worked hard to improve the efficiency of homomorphic encryption to make it more feasible in practical applications.

\subsection{Database Encoding Method}
The current efficient database encoding method~\cite{IDASH2018Andrey} for a given dataset matrix $M$ is  to flatten the matrix $M$ into a vector first and then encrypt this vector into a single ciphertext, but finally to see the resulting ciphertext as encrypting a matrix directly. This database encoding method could make full use of the HE computation and storage resources:

\begin{equation*}
 \begin{aligned}
 M = & 
\left[ \begin{array}{cccc}
 x_{10}  &   x_{11}  &  \ldots  &  x_{1d}  \\
 x_{20}  &   x_{21}  &  \ldots  &  x_{2d}  \\
 \vdots         &   \vdots         &  \ddots  &  \vdots         \\
 x_{n0}  &  x_{n1}   &  \ldots  &  x_{nd}  \\
 \end{array}
 \right]   \\
& \hspace{2.0cm}\Big\downarrow{\text{ Flattens the input dataset in a row-by-row manner }}  \\
& \left[ \begin{array}{ccccccccccccccc}
 x_{10}  &   x_{11}  &  \ldots  &  x_{1d}  & x_{20}  &   x_{21}  &  \ldots  &  x_{2d}  & \ldots  & \ldots  & \ldots  &
 x_{n0}  &  x_{n1}   &  \ldots  &  x_{nd}  \\
 \end{array}
 \right]   \\
& \hspace{2.0cm}\Big\downarrow{\text{ Encrypts the row vector }}  \\
 Enc & \left[ \begin{array}{ccccccccccccccc}
 x_{10}  &   x_{11}  &  \ldots  &  x_{1d}  & x_{20}  &   x_{21}  &  \ldots  &  x_{2d}  & \ldots  & \ldots  & \ldots  &
 x_{n0}  &  x_{n1}   &  \ldots  &  x_{nd}  \\
 \end{array}
 \right]    \\
& \hspace{2.0cm}\Big\downarrow{\text{ Seen as encrypting the dataset directly }}  \\
 Enc  & \left[ \begin{array}{cccc}
 x_{10}  &  x_{11}   &  \ldots  &  x_{1d}  \\
 x_{20}  &   x_{21}  &  \ldots  &  x_{2d}  \\
 \vdots         &   \vdots         &  \ddots  &  \vdots         \\
 x_{n0}  &   x_{n1}  &  \ldots  &  x_{nd}  \\
 \end{array}
 \right] = Enc M.  
 \end{aligned}
\end{equation*}

Based on this database encoding, two simple operations, the complete row shifting and the $incomplete$ column shifting, can be obtained by shifting the encrypted vector by two different positions $(1+d)$ and $1$, respectively:  
\begin{equation*}
 \begin{aligned}
 Enc & \left[ \begin{array}{cccc}
 x_{10}  &  x_{11}   &  \ldots  &  x_{1d}  \\
 x_{20}  &   x_{21}  &  \ldots  &  x_{2d}  \\
 \vdots         &   \vdots         &  \ddots  &  \vdots         \\
 x_{n0}  &   x_{n1}  &  \ldots  &  x_{nd}  \\
 \end{array}
 \right] 
 \xmapsto{ \text{ complete row shifiting } } 
 Enc
\left[ \begin{array}{cccc}
 x_{20}  &   x_{21}  &  \ldots  &  x_{2d}  \\
 \vdots         &   \vdots         &  \ddots  &  \vdots         \\
 x_{n0}  &   x_{n1}  &  \ldots  &  x_{nd}  \\
 x_{10}  &  x_{11}   &  \ldots  &  x_{1d}  \\
 \end{array}
 \right],  \\  \\
 Enc &  \left[ \begin{array}{cccc}
 x_{10}  &  x_{11}   &  \ldots  &  x_{1d}  \\
 x_{20}  &   x_{21}  &  \ldots  &  x_{2d}  \\
 \vdots         &   \vdots         &  \ddots  &  \vdots         \\
 x_{n0}  &   x_{n1}  &  \ldots  &  x_{nd}  \\
 \end{array}
 \right] 
 \xmapsto{ \text{ incomplete column shifiting } } 
 Enc
\left[ \begin{array}{cccc}
 x_{11}  &  \ldots  &  x_{1d}  &   x_{20}  \\
 x_{21}  &  \ldots  &  x_{2d}  &   x_{30}  \\
 \vdots         &   \vdots         &  \ddots  &  \vdots         \\
 x_{n1}  &  \ldots  &  x_{nd}  &   x_{10} \\
 \end{array}
 \right] . 
 \end{aligned}
\end{equation*}


The complete column shifting to obtain the matrix $Z^{'''}$ can also be achieved by two $\texttt{Rot}$, two $\texttt{cMult}$, and an $\texttt{Add}$.
\begin{equation*}
 \begin{aligned}
 Enc &  \left[ \begin{array}{cccc}
 x_{10}  &  x_{11}   &  \ldots  &  x_{1d}  \\
 x_{20}  &   x_{21}  &  \ldots  &  x_{2d}  \\
 \vdots         &   \vdots         &  \ddots  &  \vdots         \\
 x_{n0}  &   x_{n1}  &  \ldots  &  x_{nd}  \\
 \end{array}
 \right] 
 \xmapsto{ \text{ complete column shifiting } } 
 Enc
\left[ \begin{array}{cccc}
 x_{11}  &  \ldots  &  x_{1d}  &   x_{10}  \\
 x_{21}  &  \ldots  &  x_{2d}  &   x_{20}  \\
 \vdots         &   \vdots         &  \ddots  &  \vdots         \\
 x_{n1}  &  \ldots  &  x_{nd}  &   x_{n0} \\
 \end{array}
 \right] . 
 \end{aligned}
\end{equation*}
The same database encoding method also facilitates the development of other procedures~\cite{han2018efficient, chiang2022novel}, such as  $\texttt{SumRowVec}$ and $\texttt{SumColVec}$ to compute the sum of each row and column, respectively. 

To address the homomorphic evaluation of training 3-layer neural networks for regression and classification tasks, we introduce two procedures $\texttt{KeepOnly}$ and $\texttt{RollFill}$ :
\begin{equation*}
 \begin{aligned}
Enc & \left[ \begin{array}{ccccccccc}
 x_{10}  &   x_{11}  &  \ldots  &  x_{ij}  &  \ldots  &   x_{n0}  &  x_{n1}   &  \ldots  &  x_{nd}  \\
 \end{array}
 \right] 
 \\ & \hspace{3.0cm}  \odot    \\
& \left[ \begin{array}{ccccccccc}
~~ 0 ~~ &  ~~ 0 ~~ &  \ldots  & ~~ 1 ~~ &  \ldots  &  ~~~ 0 ~~ & ~~~ 0 ~~~  &  \ldots  & ~~ 0 ~~~ \\
 \end{array}
 \right] 
\\ & \hspace{3.06cm}  \Vert   \\
Enc & \left[ \begin{array}{ccccccccc}
~~ 0 ~~ &  ~~ 0 ~~ &  \ldots  &  x_{ij}  &  \ldots  &  ~~~ 0 ~~ & ~~~ 0 ~~~  &  \ldots  & ~~ 0 ~~~ \\
 \end{array}
 \right]    
 \\
 \\
Enc & \left[ \begin{array}{ccccccccc}
 x_{10}  &   x_{11}  &  \ldots  &  x_{ij}  &  \ldots  &   x_{n0}  &  x_{n1}   &  \ldots  &  x_{nd}  \\
 \end{array}
 \right]  \\
& \hspace{3.0cm}\Big\downarrow{\texttt{KeepOnly(i, j)}}  \\
Enc & \left[ \begin{array}{ccccccccc}
~~ 0 ~~ &  ~~ 0 ~~ &  \ldots  &  x_{ij}  &  \ldots  &  ~~~ 0 ~~ & ~~~ 0 ~~~  &  \ldots  & ~~ 0 ~~~ \\
 \end{array}
 \right]    
 \\
\\
\\
Enc & \left[ \begin{array}{ccccccccc}
~~ 0 ~~ &  ~~ 0 ~~ &  \ldots  &  x_{ij}  &  \ldots  &  ~~~ 0 ~~ & ~~~ 0 ~~~  &  \ldots  & ~~ 0 ~~~ \\
 \end{array}
 \right]  \\ 
 & \hspace{3.0cm}\Big\downarrow{\texttt{RollFill(i,j)}}  \\
Enc & \left[ \begin{array}{ccccccccc}
 x_{ij}  &   x_{ij}  &  \ldots  &  x_{ij}  &  \ldots  & ~  x_{ij}  & ~ x_{ij}   &  ~\ldots  &  x_{ij}  \\
 \end{array}
 \right] 
 \end{aligned}
\end{equation*}



\subsection{ Double Volley Revolver }
In contrast to the intricate and efficient encoding methods~\citep{kim2018matrix}, Volley Revolver~\citep{chiang2022novel} offers a straightforward, adaptable matrix-encoding approach designed specifically for privacy-preserving machine-learning tasks. Its fundamental concept, in the simplified form, involves encrypting the transpose of the second matrix before performing two matrix multiplication. 

The significance of this encoding method becomes evident in its role within privacy-preserving neural network training. Much like Chiang's observation in~\citep{chiang2022novel}, we demonstrate that Volley Revolver is indeed capable of facilitating homomorphic neural network training. This simple encoding technique aids in managing and regulating the flow of data through ciphertexts.

However, we need not limit ourselves to exclusively encrypting the transpose of the second matrix. In fact, transposing either of the two matrices can achieve the desired outcome. The option of encrypting the transpose of the first matrix is equally viable, leading to a multiplication algorithm similar to Algorithm 2 presented in~\citep{chiang2022novel}.

Furthermore, should either of the matrices prove too extensive to be encrypted into a single ciphertext, an alternative approach involves encrypting the matrices into two separate groups, labeled as Team $A$ and Team $B$, each comprising multiple ciphertexts. In this particular context, the encoding methodology referred to as  $\text{ Double Volley Revolver }$~\citep{chiang2023privacy} begins to come into play, which encompasses two distinct loops. The outer loop manages calculations involving ciphertexts from both teams, while the inner loop performs calculations on two sub-matrices encrypted by the respective ciphertexts $A_{[i]}$ and $B_{[j]}$, following the fundamental  Volley Revolver algorithm.

\subsubsection{Datasets}
We adopt three common datasets in our experiments: Boston Housing Prices, Iris and MNIST. Table \ref{tabdatasets} describes the three datasets.
\begin{table}[htbp]
\centering
\caption{Characteristics of the several datasets used in our experiments}
\label{tabdatasets}
\begin{tabular}{|c|c|c|c|c|c|c|c|c|}
\hline
Dataset &   \mysplit{No. Samples  \\ (training) }   &   \mysplit{ No. Samples  \\ (testing)  }   &  No. Features   & No. Classes  \\
\hline
Boston Housing Prices &    506   &   -   &  13   & -  \\
\hline
Iris &  150   &   -   &  4   & 3  \\
\hline
MNIST &   60,000   &   10,000   &  28$\times$28   & 10  \\
\hline
\end{tabular}
\end{table}

\section{Technical details}
%

\subsection{ 3-Layer Neural Networks }
Neural networks (NNs) are powerful machine learning algorithms engineered to capture complex nonlinear connections between input and output data, even though they may lack reasonable interpretability~\citep{chiang2023activation, hornik1989multilayer, cheng2018polynomial}. A typical NN consists of a series of layers, where iterative feedforward and backpropagation steps implement linear and non-linear transformations (activations) on input data. Each training iteration comprises a forward pass and a backward pass.




In our implementations, we utilize a 3-layer neural network comprising a single hidden layer with 12 nodes for the regression task using the Boston Housing Prices dataset. Similarly, for the classification tasks involving the Iris and MNIST datasets, we employ a 3-layer neural network featuring a single hidden layer equipped with 120 nodes.

\begin{figure}
\centering
\includegraphics[scale=1.]{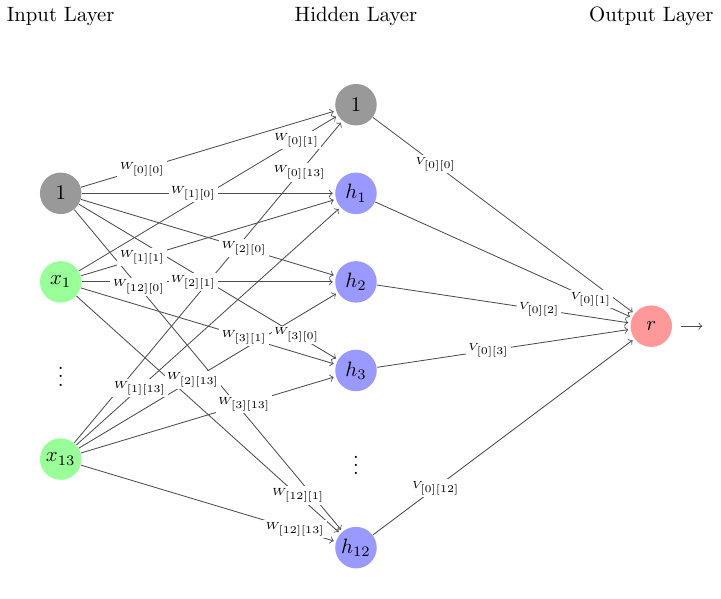}
\caption{
 The neural network architecture created for the regression task }
\label{ fig:MyHENNtrainingRegression }
\end{figure}

\begin{figure}
\centering
\includegraphics[scale=1.]{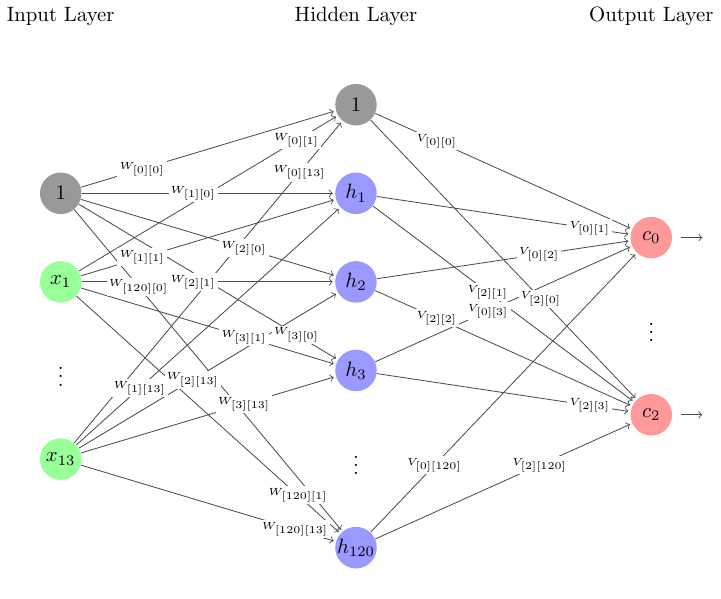}
\caption{
 The network architecture designed for the classification challenges }
\label{ fig:MyHENNtrainingClassification }
\end{figure}

Figure~\ref{ fig:MyHENNtrainingRegression } illustrates the neural network architecture created for the regression task, whereas Figure~\ref{ fig:MyHENNtrainingClassification } depicts the network architecture designed for the classification challenges.

  Given the matrix $X \in \mathbb{R}^{n \times (1 + d)}$, the column vector $Y \in \mathbb{N}^{n \times 1}$, the matrix $\bar{Y} \in \mathbb{R}^{n \times c}$, the matrix $W \in \mathbb{R}^{m \times (1 + d)}$, and the matrix $V \in \mathbb{R}^{c \times (1 + m)}$, we interpret these entities as follows: $X$ represents the dataset, $Y$ denotes the class labels in column vector format, $\bar{Y}$ signifies the one-hot encoding of the class labels, $W$ stands for the weight matrix connecting the first two layers of the neural network, and $V$ symbolizes the weight matrix linking the last two layers. Here, $m$ is the number of nodes in the hidden layer.
  
\begin{align*}
  X &= 
 \begin{bmatrix}
 \textsl x_{[1]}      \\
 \textsl x_{[2]}      \\
 \vdots          \\
 \textsl x_{[n]}      \\
 \end{bmatrix}
 = 
 \begin{bmatrix}
 1    &   x_{[1][1]}   &  \cdots  & x_{[1][d]}   \\
 1    &   x_{[2][1]}   &  \cdots  & x_{[2][d]}   \\
 \vdots    &   \vdots   &  \ddots  & \vdots   \\
 1    &   x_{[n][1]}   &  \cdots  & x_{[n][d]}   \\
 \end{bmatrix}
  = 
 \begin{bmatrix}
 x_{[1][0]}    &   x_{[1][1]}   &  \cdots  & x_{[1][d]}   \\
 x_{[2][0]}    &   x_{[2][1]}   &  \cdots  & x_{[2][d]}   \\
 \vdots    &   \vdots   &  \ddots  & \vdots   \\
 x_{[n][0]}    &   x_{[n][1]}   &  \cdots  & x_{[n][d]}   \\
 \end{bmatrix},   \\   
 Y &=  
 \begin{bmatrix}
 y_{1}     \\
 y_{2}     \\
 \vdots         \\
 y_{n}     \\
 \end{bmatrix}
\xmapsto{ \text{one-hot encoding} } 
\textsl Y =
 \begin{bmatrix}
 { \textsl{y}_{[1]} }     \\
 { \textsl{y}_{[2]} }     \\
 \vdots         \\
 { \textsl{y}_{[n]} }     \\
 \end{bmatrix} 
 =
 \begin{bmatrix}
 y_{[1][1]}    &   y_{[1][2]}   &  \cdots  & y_{[1][c]}   \\
 y_{[2][1]}    &   y_{[2][2]}   &  \cdots  & y_{[2][c]}   \\
 \vdots    &   \vdots   &  \ddots  & \vdots   \\
 y_{[n][1]}    &   y_{[n][2]}   &  \cdots  & y_{[n][c]}   \\
 \end{bmatrix}, \\
 W &=  
 \begin{bmatrix}
 \textsl w_{[1]}     \\
 \textsl w_{[2]}     \\
 \vdots         \\
 \textsl w_{[m]}     \\
 \end{bmatrix}  =
 \begin{bmatrix}
 w_{[1][0]}    &   w_{[1][1]}   &  \cdots  & w_{[1][d]}   \\
 w_{[2][0]}    &   w_{[2][1]}   &  \cdots  & w_{[2][d]}   \\
 \vdots    &   \vdots   &  \ddots  & \vdots   \\
 w_{[m][0]}    &   w_{[m][1]}   &  \cdots  & w_{[m][d]}   \\
 \end{bmatrix},  \\
 V &=  
 \begin{bmatrix}
 \textsl v_{[1]}     \\
 \textsl v_{[2]}     \\
 \vdots         \\
 \textsl v_{[c]}     \\
 \end{bmatrix}  =
 \begin{bmatrix}
 v_{[1][0]}    &   v_{[1][1]}   &  \cdots  & v_{[1][m]}   \\
 v_{[2][0]}    &   v_{[2][1]}   &  \cdots  & v_{[2][m]}   \\
 \vdots    &   \vdots   &  \ddots  & \vdots   \\
 v_{[c][0]}    &   v_{[c][1]}   &  \cdots  & v_{[c][m]}   \\
 \end{bmatrix}.
\end{align*} 

NN training conssits of two stages: Forward inference and Backward training.

\paragraph{Forward inference}
\begin{enumerate}[start=1,label={\texttt{Step} \arabic*:}]
	\item \begin{align*}
  X \times W^{\intercal} &= 
 \begin{bmatrix}
 z_{[1][1]}    &   z_{[1][2]}   &  \cdots  & z_{[1][m]}   \\
 z_{[2][1]}    &   z_{[2][2]}   &  \cdots  & z_{[2][m]}   \\
 \vdots    &   \vdots   &  \ddots  & \vdots   \\
 z_{[n][1]}    &   z_{[n][2]}   &  \cdots  & z_{[n][m]}   \\
 \end{bmatrix}  =   Z_0 ,  \\
  \end{align*}
  \item  we select the square function as the activation function denoted as $\phi ( x ) = x^2 $  
   \begin{align*}
 Z_0 & \xmapsto{ \phi ( Z ) }  Z_{1} =
  \begin{bmatrix}
 \phi( z_{[1][1]} )   &   \phi( z_{[1][2]} )   &  \cdots  & \phi( z_{[1][m]} )   \\
 \phi( z_{[2][1]} )    &   \phi( z_{[2][2]} )   &  \cdots  & \phi( z_{[2][m]} )   \\
 \vdots    &   \vdots   &  \ddots  & \vdots   \\
 \phi( z_{[n][1]} )    &   \phi( z_{[n][2]} )   &  \cdots  & \phi( z_{[n][m]} )   \\
 \end{bmatrix},   \\
  \end{align*} 
  \item  \begin{align*}
 Z_{1} & \xmapsto{  }  Z =
  \begin{bmatrix}
 1      &      \phi( z_{[1][1]} )   &   \phi( z_{[1][2]} )   &  \cdots  & \phi( z_{[1][m]} )   \\
 1      &      \phi( z_{[2][1]} )    &   \phi( z_{[2][2]} )   &  \cdots  & \phi( z_{[2][m]} )   \\
 \vdots      &      \vdots    &   \vdots   &  \ddots  & \vdots   \\
 1      &      \phi( z_{[n][1]} )    &   \phi( z_{[n][2]} )   &  \cdots  & \phi( z_{[n][m]} )   \\
 \end{bmatrix},   \\
  \end{align*} 
  \item  \begin{align*}
 Z \times V^{\intercal} &= \bar Y = 
 \begin{bmatrix}
 \bar y_{[1][1]}    &   \bar y_{[1][2]}   &  \cdots  & \bar y_{[1][c]}   \\
 \bar y_{[2][1]}    &   \bar y_{[2][2]}   &  \cdots  & \bar y_{[2][c]}   \\
 \vdots    &   \vdots   &  \ddots  & \vdots   \\
 \bar y_{[n][1]}    &   \bar y_{[n][2]}   &  \cdots  & \bar y_{[n][c]}   \\
 \end{bmatrix}, \\
  \end{align*} 
\end{enumerate}

\paragraph{Backward training}
$\nabla W = (S \times \bar V \odot Z^{'})^{\intercal} \times X$ , $\nabla V = S^{\intercal} \times Z$ where $\bar V $ is
\begin{align*}
\bar V &=  
 \begin{bmatrix}
 v_{[1][1]}    &   v_{[1][2]}   &  \cdots  & v_{[1][m]}   \\
 v_{[2][1]}    &   v_{[2][2]}   &  \cdots  & v_{[2][m]}   \\
 \vdots    &   \vdots   &  \ddots  & \vdots   \\
 v_{[c][1]}    &   v_{[c][2]}   &  \cdots  & v_{[c][m]}   \\
 \end{bmatrix},  \\   
 \end{align*}  
 and $ Z^{'}$ is obtained from
 \begin{align*}
 Z_{1} & \xmapsto{ \nabla }  Z^{'} =
  \begin{bmatrix}
 \phi^{'}( z_{[1][1]} )   &   \phi^{'}( z_{[1][2]} )   &  \cdots  & \phi^{'}( z_{[1][m]} )   \\
 \phi^{'}( z_{[2][1]} )    &   \phi^{'}( z_{[2][2]} )   &  \cdots  & \phi^{'}( z_{[2][m]} )   \\
 \vdots    &   \vdots   &  \ddots  & \vdots   \\
 \phi^{'}( z_{[n][1]} )    &   \phi^{'}( z_{[n][2]} )   &  \cdots  & \phi^{'}( z_{[n][m]} )   \\
 \end{bmatrix}.
\end{align*} 
Finally, various first-order gradient descent algorithms with a certain learning rate scheme can be used to modify the NN parameters $W$ and $V$. In our work, we just adopt the raw gradient descent with a fixed learning rate $\eta$: $W = W - \eta \cdot \nabla W$ and $V = V - \eta \cdot \nabla V$ .

\subsection{ Approxiting Softmax Function }
The conventional training approach for classification involves utilizing the log-likelihood loss function, which incorporates the Softmax function:
\begin{align*}
\bar Y    & \xmapsto{ Softmax } S_0 = \\
 &\begin{bmatrix}
 \exp( \bar y_{[1][1]} ) / \sum_{i=1}^{c} \exp( \bar y_{[1][i]} )     &   \exp( \bar y_{[1][2]} ) / \sum_{i=1}^{c} \exp( \bar y_{[1][i]} )   &  \cdots  & \exp( \bar y_{[1][c]} ) / \sum_{i=1}^{c} \exp( \bar y_{[1][i]} )   \\
 \exp( \bar y_{[2][1]} ) / \sum_{i=1}^{c} \exp( \bar y_{[2][i]} )    &   \exp( \bar y_{[2][2]} ) / \sum_{i=1}^{c} \exp( \bar y_{[2][i]} )   &  \cdots  & \exp( \bar y_{[2][c]} ) / \sum_{i=1}^{c} \exp( \bar y_{[2][i]} )   \\
 \vdots    &   \vdots   &  \ddots  & \vdots   \\
 \exp( \bar y_{[n][1]} ) / \sum_{i=1}^{c} \exp( \bar y_{[n][i]} )    &   \exp( \bar y_{[n][2]} ) / \sum_{i=1}^{c} \exp( \bar y_{[n][i]} )   &  \cdots  & \exp( \bar y_{[n][c]} ) / \sum_{i=1}^{c} \exp( \bar y_{[n][i]} )   \\
 \end{bmatrix}, \\
 S_0  & \xmapsto{ \mathop{\arg\max} } output = 
 \begin{bmatrix}
 \mathop{\arg\max}_{j}  \exp( \bar y_{[1][j]} ) / \sum_{i=1}^{c} \exp( \bar y_{[1][i]} )    \\
 \mathop{\arg\max}_{j}  \exp( \bar y_{[2][j]} ) / \sum_{i=1}^{c} \exp( \bar y_{[1][i]} )    \\
 \vdots      \\
 \mathop{\arg\max}_{j}  \exp( \bar y_{[n][j]} ) / \sum_{i=1}^{c} \exp( \bar y_{[1][i]} )    \\
 \end{bmatrix}. \\
\end{align*}

Given the presence of inherent uncertainties, it could be challenging to attain an acceptably-usable polynomial approximation of the Softmax function within the context of privacy-preserving computations.
To tackle this challenge, Chiang~\citep{chiang2023privacy} apply the mathematical strategy: transforming a complex problem into a simpler counterpart. Considering this, rather than attempting a direct approximation of the Softmax function, they shift their focus towards approximating the Sigmoid function within the encrypted domain. In doing so, they have developed a novel loss function named $SLE$ (Squared Likelihood  Error), which solely relies on the Sigmoid function, rather than using the Softmax function.

\subsection{Squared Likelihood  Error}

The SLE loss function is expected to yield the following neural network output:

\begin{align*}
\bar Y    & \xmapsto{ SLE_1 } S_1 = \\
 &\begin{bmatrix}
 Sigmoid ( \bar y_{[1][1]} )     &   Sigmoid ( \bar y_{[1][2]} )   &  \cdots  & Sigmoid ( \bar y_{[1][c]} )    \\
 Sigmoid ( \bar y_{[2][1]} )   &   Sigmoid ( \bar y_{[2][2]} )    &  \cdots  & Sigmoid ( \bar y_{[2][c]} )   \\
 \vdots    &   \vdots   &  \ddots  & \vdots   \\
 Sigmoid ( \bar y_{[n][1]} )     &   Sigmoid ( \bar y_{[n][2]} )  &  \cdots  & Sigmoid ( \bar y_{[n][c]} )   \\
 \end{bmatrix}, \\
 S_1  & \xmapsto{ \mathop{\arg\max} } output = 
 \begin{bmatrix}
 \mathop{\arg\max}_{j}  Sigmoid ( \bar y_{[1][j]} )     \\
 \mathop{\arg\max}_{j}  Sigmoid ( \bar y_{[2][j]} )     \\
 \vdots      \\
 \mathop{\arg\max}_{j}  Sigmoid ( \bar y_{[n][j]} )     \\
 \end{bmatrix}, \\
\end{align*} 
and thus has the $S$:
\begin{align*} 
 S &= 
 \begin{bmatrix}
 s_{[1][1]}    &   s_{[1][2]}   &  \cdots  & s_{[1][c]}   \\
 s_{[2][1]}    &   s_{[2][2]}   &  \cdots  & s_{[2][c]}   \\
 \vdots    &   \vdots   &  \ddots  & \vdots   \\
 s_{[n][1]}    &   s_{[n][2]}   &  \cdots  & s_{[n][c]}   \\
 \end{bmatrix} \\  
 & = 
 \begin{bmatrix}
 1  - Sigmoid(\bar y_{[1][1]} ) - y_{[1][1]}    &   1  - Sigmoid(\bar y_{[1][2]} ) - y_{[1][2]}   &  \cdots  & 1  - Sigmoid(\bar y_{[1][c]} ) - y_{[1][c]}   \\
 1  - Sigmoid(\bar y_{[2][1]} ) - y_{[2][1]}    &   1  - Sigmoid(\bar y_{[2][2]} ) - y_{[2][2]}   &  \cdots  & 1  - Sigmoid(\bar y_{[2][c]} ) - y_{[2][c]}   \\
 \vdots    &   \vdots   &  \ddots  & \vdots   \\
 1  - Sigmoid(\bar y_{[n][1]} ) - y_{[n][1]}    &   1  - Sigmoid(\bar y_{[n][2]} ) - y_{[n][2]}   &  \cdots  & 1  - Sigmoid(\bar y_{[n][c]} ) - y_{[n][c]}   \\
 \end{bmatrix}.
\end{align*} 

The SLE loss function, denoted as $L$, along with its logarithmic differential $\mathrm{d}  \ln L$, is defined as follows: 
$$L = \prod _{i=1}^n  \prod _{j=1}^{c} ( Sigmoid( \bar y_{[i][j]} ) -  y_{[i][j]}  )^2      \xmapsto{ \text{\ \ \ \ \ \ } }      \ln {L} = \sum _{i=1}^n  \sum _{j=1}^{c} \ln | Sigmoid( \bar y_{[i][j]} ) - y_{[i][j]}  |   , $$ and  
 $$      \mathrm{d}  \ln {L} = \sum _{i=1}^n  \sum _{j=1}^{c} ( 1 - Sigmoid(\bar y_{[i][j]}) - y_{[i][j]}) \mathrm{d} \bar y_{[i][j]}  .           $$

\subsubsection{First Variant of SLE}
Although SLE demonstrates success in multiclass logistic regression, as evidenced in~\citep{chiang2023privacy}, it fails to yield effective results in neural networks with hidden layers. This could be attributed to a tendency to easily converge towards local minima. 

We propose a variant of SLE that can address this issue still using only the Sigmoid function:
 $\texttt{1st variant of SLE}$:  $$L_1 = \sum _{i=1}^n  \sum _{j=1}^{c} ( Sigmoid( \bar y_{[i][j]} ) -  y_{[i][j]}  )^2   . $$ 
 
Its differential $\mathrm{d}  L_1$, is defined as follows: 
\begin{align*}
 \mathrm{d} L_1 &=  \sum _{i=1}^n  \sum _{j=1}^{c} \mathrm{d} ( Sigmoid( \bar y_{[i][j]} ) -  y_{[i][j]}  )^2          \\
  &=  \sum _{i=1}^n  \sum _{j=1}^{c}  2 \cdot ( Sigmoid( \bar y_{[i][j]} ) -  y_{[i][j]}  ) \cdot Sigmoid( \bar y_{[i][j]} ) \cdot (1 - Sigmoid( \bar y_{[i][j]} ))  \mathrm{d} \bar y_{[i][j]}  ,     \end{align*}
 and a simplified approximation of $\mathrm{d} L_1$ also demonstrates remarkable performance while requiring fewer calculations:
$$ \mathrm{d} L_1 =  \sum _{i=1}^n  \sum _{j=1}^{c}  2 \cdot ( Sigmoid( \bar y_{[i][j]} ) -  y_{[i][j]}  ) \cdot 0.25  \mathrm{d} \bar y_{[i][j]}  .
$$

 
For this variant of the $SLE$ loss function, we can formulate the expression for $S$ as follows:
\begin{align*} 
 S &= 
  \begin{bmatrix}
 s_{[1][1]}    &   s_{[1][2]}   &  \cdots  & s_{[1][c]}   \\
 s_{[2][1]}    &   s_{[2][2]}   &  \cdots  & s_{[2][c]}   \\
 \vdots    &   \vdots   &  \ddots  & \vdots   \\
 s_{[n][1]}    &   s_{[n][2]}   &  \cdots  & s_{[n][c]}   \\
 \end{bmatrix} \\
 &=
 \begin{bmatrix}
 2 \cdot ( Sigmoid( \bar y_{[1][1]} ) -  y_{[1][1]}  ) \cdot 0.25      &  \cdots  & 2 \cdot ( Sigmoid( \bar y_{[1][c]} ) -  y_{[1][c]}  ) \cdot 0.25   \\
 2 \cdot ( Sigmoid( \bar y_{[2][1]} ) -  y_{[2][1]}  ) \cdot 0.25    &    \cdots  & 2 \cdot ( Sigmoid( \bar y_{[2][c]} ) -  y_{[2][c]}  ) \cdot 0.25   \\
 \vdots    &    \ddots  & \vdots   \\
 2 \cdot ( Sigmoid( \bar y_{[n][1]} ) -  y_{[n][1]}  ) \cdot 0.25    &     \cdots  & 2 \cdot ( Sigmoid( \bar y_{[n][c]} ) -  y_{[n][c]}  ) \cdot 0.25   \\
 \end{bmatrix}.
\end{align*} 

\subsubsection{Second Variant of SLE}

The aforementioned initial variant of SLE with the expression $L_1$ demonstrates consistent, robust, and high performance when applied to neural networks with a single hidden layer. However, the input values to the Sigmoid function tend to be large. For instance, when implementing $SLEL_1$ on the MNIST dataset, the Sigmoid function often encounters input values wider than the range of $[-25, +25]$. Even a minor discrepancy from the Sigmoid function itself can adversely affect the performance of the first variant of SLE. While existing HE techniques can achieve near-perfect approximations of the Sigmoid function within such a range, it requires a significant amount of homomorphic computations in the encrypted domain. This drawback results in a less-than-ideal practical solution.

We introduce a second novel variant of SLE that can address the limitations of the first version. This second variant bears a strong resemblance to the Mean Squared Error (MSE) loss function and solely relies on the Sigmoid function:

$\texttt{2nd variant of SLE}$:  $$L_2 = \sum _{i=1}^n  \sum _{j=1}^{c} ( \bar y_{[i][j]}  -  y_{[i][j]}  )^2      . $$ 
 
Its differential $\mathrm{d}  L_2$, is defined as follows: 
  $$      \mathrm{d}  {L_2} = \sum _{i=1}^n  \sum _{j=1}^{c} 2 \cdot ( \bar y_{[i][j]}  -  y_{[i][j]}  ) \cdot \mathrm{d} \bar y_{[i][j]}     .      $$

For this particular variant of the $SLE$ loss function, we can deduce the expression for $S$:
\begin{align*} 
 S &= 
  \begin{bmatrix}
 s_{[1][1]}    &   s_{[1][2]}   &  \cdots  & s_{[1][c]}   \\
 s_{[2][1]}    &   s_{[2][2]}   &  \cdots  & s_{[2][c]}   \\
 \vdots    &   \vdots   &  \ddots  & \vdots   \\
 s_{[n][1]}    &   s_{[n][2]}   &  \cdots  & s_{[n][c]}   \\
 \end{bmatrix}  \\
 &=
 \begin{bmatrix}
 2 \cdot ( \bar y_{[1][1]}  -  y_{[1][1]}  )    &   2 \cdot ( \bar y_{[1][2]}  -  y_{[1][2]}  )    &  \cdots  & 2 \cdot ( \bar y_{[1][c]}  -  y_{[1][c]}  )   \\
 2 \cdot (  \bar y_{[2][1]}  -  y_{[2][1]}  )     &   2 \cdot ( \bar y_{[2][2]}  -  y_{[2][2]}  )    &  \cdots  & 2 \cdot (  \bar y_{[2][c]}  -  y_{[2][c]}  )   \\
 \vdots    &   \vdots   &  \ddots  & \vdots   \\
 2 \cdot (  \bar y_{[n][1]}  -  y_{[n][1]}  )     &   2 \cdot (  \bar y_{[n][2]}  -  y_{[n][2]}  )    &  \cdots  & 2 \cdot (  \bar y_{[n][c]}  -  y_{[n][c]}  )    \\
 \end{bmatrix}.
\end{align*} 

\begin{figure}[htp]
\centering
\captionsetup[subfigure]{justification=centering}

\subfloat[MNIST Testing Precision]{\includegraphics[width=7cm]{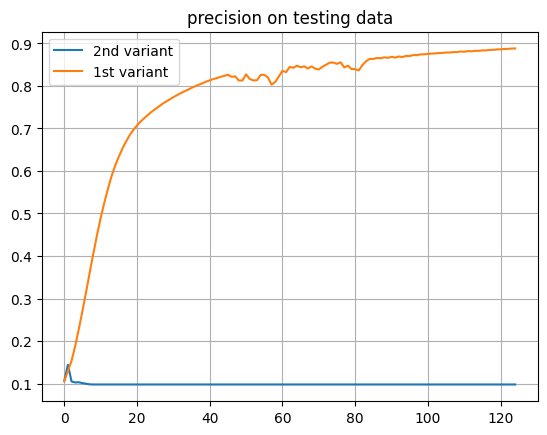}\label{fig:subfig01}}
\subfloat[MNIST Training Precision]{\includegraphics[width=7cm]{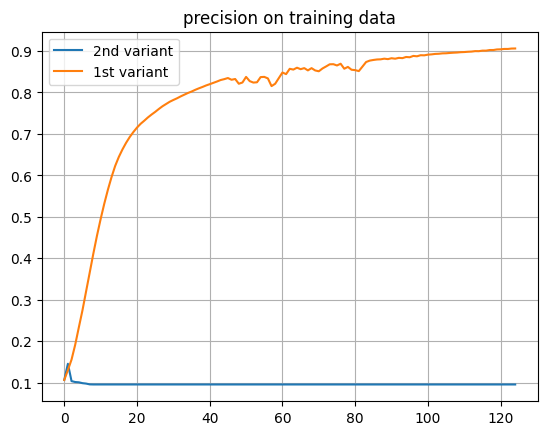}\label{fig:subfig02}}

\subfloat[MNIST Testing  Loss function]{\includegraphics[width=7cm]{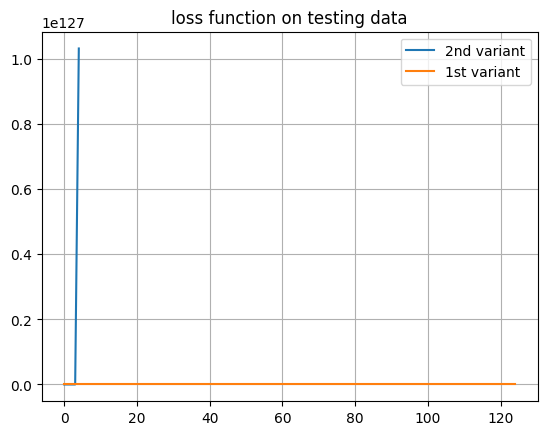}\label{fig:subfig03}}
\subfloat[MNIST Training  Loss function]{\includegraphics[width=7cm]{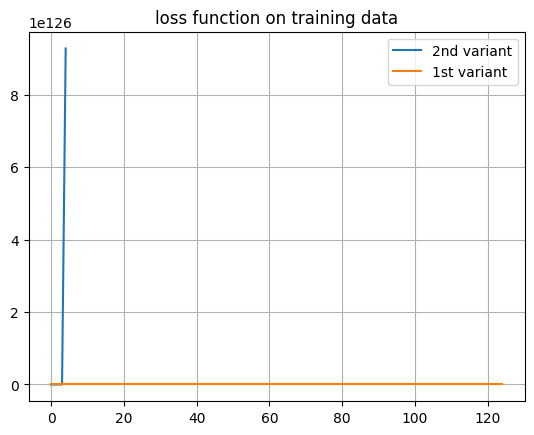}\label{fig:subfig04}}

\caption{Training and testing experimental results for the two variants of SLE are provided using the same learning rate $0.12$ }
\label{fig1}
\end{figure}

\begin{figure}[htp]
\centering
\captionsetup[subfigure]{justification=centering}

\subfloat[MNIST Testing Precision]{\includegraphics[width=7cm]{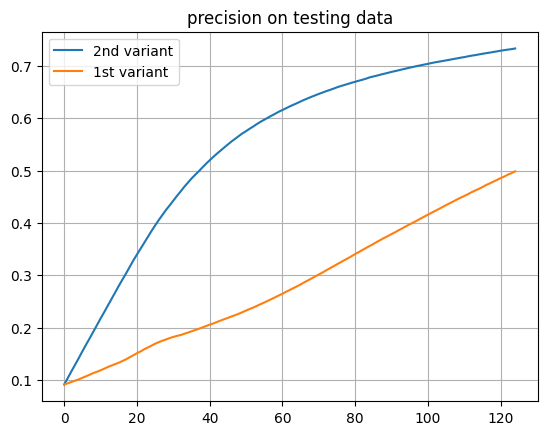}\label{fig:subfig01}}
\subfloat[MNIST Training Precision]{\includegraphics[width=7cm]{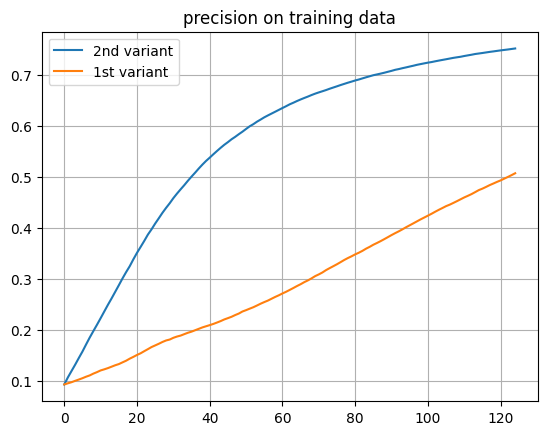}\label{fig:subfig02}}

\subfloat[MNIST Testing  Loss function]{\includegraphics[width=7cm]{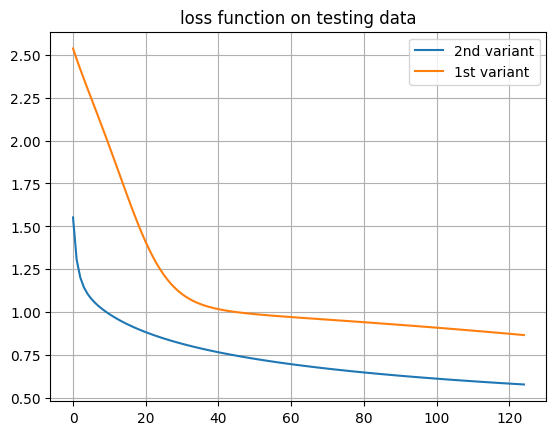}\label{fig:subfig03}}
\subfloat[MNIST Training  Loss function]{\includegraphics[width=7cm]{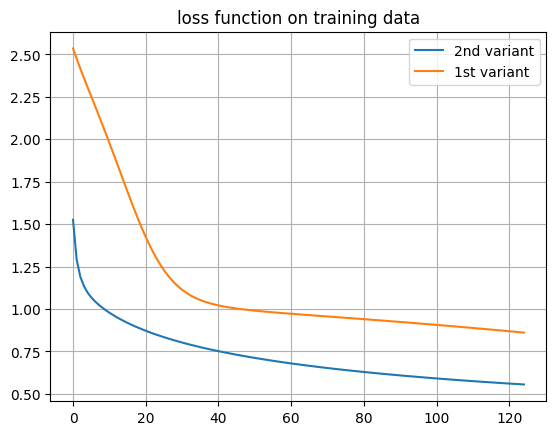}\label{fig:subfig04}}

\caption{Training and testing experimental results for the two variants of SLE are provided using the same learning rate $0.01$}
\label{fig2}
\end{figure}

\paragraph{$\texttt{Performance Evaluation}$}
The two variants of SLE possess their individual advantages and disadvantages. We employ the first $5,000$ MNIST training images to train the NN model with 120 hidden nodes, while utilizing the complete MNIST testing dataset to evaluate the generated performance of the resulting model. This test is repeated 12 times, utilizing two distinct learning rates for the two loss function variants. The average performances in terms of loss and accuracy are presented in Figure~\ref{fig1} and Figure~\ref{fig2}.

\section{Homomorphic NN Training}

For simplicity, we utilize the extreme case of the "Double Volley Revolver" methodology to illustrate the feasibility of homomorphic training for a 3-layer neural network (NN) in both regression and classification tasks. In this scenario, we make the assumption that each individual ciphertext can only encrypt a single row vector such as $\textsl x_{[i]}$, $\textsl{y}{[j]}$, $\textsl w{[k]}$, or $\textsl v_{[g]}$:

 \begin{align*} 
   X & = 
 \begin{bmatrix}
 x_{[1][0]}    &   x_{[1][1]}   &  \cdots  & x_{[1][d]}   \\
 x_{[2][0]}    &   x_{[2][1]}   &  \cdots  & x_{[2][d]}   \\
 \vdots    &   \vdots   &  \ddots  & \vdots   \\
 x_{[n][0]}    &   x_{[n][1]}   &  \cdots  & x_{[n][d]}   \\
 \end{bmatrix} 
 \longrightarrow 
 \begin{bmatrix}
  \texttt{Enc}
 \begin{mbmatrix}
  x_{[1][0]}    &   x_{[1][1]}    &  \ldots  & x_{[1][d]}    &   0   &  \ldots  & 0 
 \end{mbmatrix}      \\
  \texttt{Enc}
 \begin{mbmatrix}
  x_{[2][0]}    &   x_{[2][1]}    &  \ldots  & x_{[2][d]}    &   0   &  \ldots  & 0
 \end{mbmatrix}     \\
 \vdots        \\
  \texttt{Enc}
 \begin{mbmatrix}
  x_{[n][0]}    &   x_{[n][1]}    &  \ldots  & x_{[n][d]}     &   0   &  \ldots  & 0 
 \end{mbmatrix}   \\
 \end{bmatrix},  
\end{align*}

 \begin{align*} 
 Y &=  
 \begin{bmatrix}
 y_{[1][1]}    &   y_{[1][2]}   &  \cdots  & y_{[1][c]}   \\
 y_{[2][1]}    &   y_{[2][2]}   &  \cdots  & y_{[2][c]}   \\
 \vdots    &   \vdots   &  \ddots  & \vdots   \\
 y_{[n][1]}    &   y_{[n][2]}   &  \cdots  & y_{[n][c]}   \\
 \end{bmatrix}
  \longrightarrow 
 \begin{bmatrix}
  \texttt{Enc}
 \begin{mbmatrix}
  y_{[1][1]}    &   y_{[1][2]}    &  \ldots  & y_{[1][c]}    &   0   &  \ldots  & 0 
 \end{mbmatrix}      \\
  \texttt{Enc}
 \begin{mbmatrix}
  y_{[2][1]}    &   y_{[2][2]}     &  \ldots  & y_{[2][c]}    &   0   &  \ldots  & 0
 \end{mbmatrix}     \\
 \vdots        \\
  \texttt{Enc}
 \begin{mbmatrix}
  y_{[n][1]}     &  y_{[n][2]}    &  \ldots  & y_{[n][c]}    &   0   &  \ldots  & 0 
 \end{mbmatrix}   \\
 \end{bmatrix},  
\end{align*}

 \begin{align*} 
 W &=  
 \begin{bmatrix}
 w_{[1][0]}    &   w_{[1][1]}   &  \cdots  & w_{[1][d]}   \\
 w_{[2][0]}    &   w_{[2][1]}   &  \cdots  & w_{[2][d]}   \\
 \vdots    &   \vdots   &  \ddots  & \vdots   \\
 w_{[m][0]}    &   w_{[m][1]}   &  \cdots  & w_{[m][d]}   \\
 \end{bmatrix}
  \longrightarrow 
 \begin{bmatrix}
  \texttt{Enc}
 \begin{mbmatrix}
  w_{[1][0]}    &   w_{[1][1]}    &  \ldots  & w_{[1][d]}    &   0   &  \ldots  & 0 
 \end{mbmatrix}      \\
  \texttt{Enc}
 \begin{mbmatrix}
  w_{[2][0]}    &   w_{[2][1]}    &  \ldots  & w_{[2][d]}    &   0   &  \ldots  & 0
 \end{mbmatrix}     \\
 \vdots        \\
  \texttt{Enc}
 \begin{mbmatrix}
  w_{[m][0]}      &  w_{[m][1]}    &  \ldots  & w_{[m][d]}    &   0   &  \ldots  & 0 
 \end{mbmatrix}   \\
 \end{bmatrix},  
\end{align*}

 \begin{align*} 
 V &=  
 \begin{bmatrix}
 v_{[1][0]}    &   v_{[1][1]}   &  \cdots  & v_{[1][m]}   \\
 v_{[2][0]}    &   v_{[2][1]}   &  \cdots  & v_{[2][m]}   \\
 \vdots    &   \vdots   &  \ddots  & \vdots   \\
 v_{[c][0]}    &   v_{[c][1]}   &  \cdots  & v_{[c][m]}   \\
 \end{bmatrix}
  \longrightarrow 
 \begin{bmatrix}
  \texttt{Enc}
 \begin{mbmatrix}
  v_{[1][0]}    &   v_{[1][1]}     &  \ldots  &  v_{[1][m]}    &   0   &  \ldots  & 0 
 \end{mbmatrix}      \\
  \texttt{Enc}
 \begin{mbmatrix}
  v_{[2][0]}    &   v_{[2][1]}    &  \ldots  & v_{[2][m]}    &   0   &  \ldots  & 0
 \end{mbmatrix}     \\
 \vdots        \\
  \texttt{Enc}
 \begin{mbmatrix}
  v_{[c][0]}     &  v_{[c][1]}    &  \ldots  & v_{[c][m]}    &   0   &  \ldots  & 0 
 \end{mbmatrix}   \\
 \end{bmatrix}.  
\end{align*}

By using HE operations merely  we can obtain the ciphertexts that encrypt each row of the matrices $S$, $\bar V$, $Z^{'}$  and $Z$ : 

 \begin{align*} 
 S &= 
  \begin{bmatrix}
 s_{[1][1]}    &   s_{[1][2]}   &  \cdots  & s_{[1][c]}   \\
 s_{[2][1]}    &   s_{[2][2]}   &  \cdots  & s_{[2][c]}   \\
 \vdots    &   \vdots   &  \ddots  & \vdots   \\
 s_{[n][1]}    &   s_{[n][2]}   &  \cdots  & s_{[n][c]}   \\
 \end{bmatrix} 
   \longrightarrow 
 \begin{bmatrix}
  \texttt{Enc}
 \begin{mbmatrix}
  s_{[1][1]}     &   s_{[1][2]}     &  \ldots  &  s_{[1][c]}     &   0   &  \ldots  & 0 
 \end{mbmatrix}      \\
  \texttt{Enc}
 \begin{mbmatrix}
  s_{[2][1]}    &   s_{[2][2]}    &  \ldots  & s_{[2][c]}    &   0   &  \ldots  & 0
 \end{mbmatrix}     \\
 \vdots        \\
  \texttt{Enc}
 \begin{mbmatrix}
  s_{[n][1]}     &  s_{[n][2]}    &  \ldots  & s_{[n][c]}   &   0   &  \ldots  & 0 
 \end{mbmatrix}   \\
 \end{bmatrix},  
\end{align*}

 \begin{align*} 
\bar V &=  
 \begin{bmatrix}
 v_{[1][1]}    &   v_{[1][2]}   &  \cdots  & v_{[1][m]}   \\
 v_{[2][1]}    &   v_{[2][2]}   &  \cdots  & v_{[2][m]}   \\
 \vdots    &   \vdots   &  \ddots  & \vdots   \\
 v_{[c][1]}    &   v_{[c][2]}   &  \cdots  & v_{[c][m]}   \\
 \end{bmatrix}
   \longrightarrow 
 \begin{bmatrix}
  \texttt{Enc}
 \begin{mbmatrix}
  v_{[1][1]}     &   v_{[1][2]}     &  \ldots  &  v_{[1][m]}     &   0   &  \ldots  & 0 
 \end{mbmatrix}      \\
  \texttt{Enc}
 \begin{mbmatrix}
  v_{[2][1]}    &   v_{[2][2]}    &  \ldots  & v_{[2][m]}    &   0   &  \ldots  & 0
 \end{mbmatrix}     \\
 \vdots        \\
  \texttt{Enc}
 \begin{mbmatrix}
  v_{[c][1]}     &  v_{[c][2]}    &  \ldots  & v_{[c][m]}   &   0   &  \ldots  & 0 
 \end{mbmatrix}   \\
 \end{bmatrix},  
\end{align*}

 \begin{align*} 
Z^{'} & =
  \begin{bmatrix}
 \phi^{'}( z_{[1][1]} )   &   \phi^{'}( z_{[1][2]} )   &  \cdots  & \phi^{'}( z_{[1][m]} )   \\
 \phi^{'}( z_{[2][1]} )    &   \phi^{'}( z_{[2][2]} )   &  \cdots  & \phi^{'}( z_{[2][m]} )   \\
 \vdots    &   \vdots   &  \ddots  & \vdots   \\
 \phi^{'}( z_{[n][1]} )    &   \phi^{'}( z_{[n][2]} )   &  \cdots  & \phi^{'}( z_{[n][m]} )   \\
 \end{bmatrix}  \\
  & \longrightarrow 
  \begin{bmatrix}
  \texttt{Enc}
 \begin{mbmatrix}
  \phi^{'}( z_{[1][1]} )     &   \phi^{'}( z_{[1][2]} )     &  \ldots  &  \phi^{'}( z_{[1][m]} )     &   0   &  \ldots  & 0 
 \end{mbmatrix}      \\
  \texttt{Enc}
 \begin{mbmatrix}
  \phi^{'}( z_{[2][1]} )    &   \phi^{'}( z_{[2][2]} )    &  \ldots  & \phi^{'}( z_{[2][m]} )    &   0   &  \ldots  & 0
 \end{mbmatrix}     \\
 \vdots        \\
  \texttt{Enc}
 \begin{mbmatrix}
  \phi^{'}( z_{[n][1]} )     &  \phi^{'}( z_{[n][2]} )   &  \ldots  & \phi^{'}( z_{[n][m]} )   &   0   &  \ldots  & 0 
 \end{mbmatrix}   \\
 \end{bmatrix},  
\\
Z & =
  \begin{bmatrix}
 1      &      \phi( z_{[1][1]} )   &   \phi( z_{[1][2]} )   &  \cdots  & \phi( z_{[1][m]} )   \\
 1      &      \phi( z_{[2][1]} )    &   \phi( z_{[2][2]} )   &  \cdots  & \phi( z_{[2][m]} )   \\
 \vdots      &      \vdots    &   \vdots   &  \ddots  & \vdots   \\
 1      &      \phi( z_{[n][1]} )    &   \phi( z_{[n][2]} )   &  \cdots  & \phi( z_{[n][m]} )   \\
 \end{bmatrix}   \\
 &  \longrightarrow 
 \begin{bmatrix}
  \texttt{Enc}
 \begin{mbmatrix}
  1      &      \phi( z_{[1][1]} )   &   \phi( z_{[1][2]} )   &  \cdots  & \phi( z_{[1][m]} )     &   0   &  \ldots  & 0 
 \end{mbmatrix}      \\
  \texttt{Enc}
 \begin{mbmatrix}
  1      &      \phi( z_{[2][1]} )    &   \phi( z_{[2][2]} )   &  \cdots  & \phi( z_{[2][m]} )    &   0   &  \ldots  & 0
 \end{mbmatrix}     \\
 \vdots        \\
  \texttt{Enc}
 \begin{mbmatrix}
 1      &      \phi( z_{[n][1]} )    &   \phi( z_{[n][2]} )   &  \cdots  & \phi( z_{[n][m]} )    &   0   &  \ldots  & 0 
 \end{mbmatrix}   \\
 \end{bmatrix},  
\end{align*}

\subsection{ For Classification Problems }

We can first compute the gradients of $\textsl w_{[k]}$ and $\textsl v_{[k]}$ in the encrypted environment and then use the regular gradient descent method to update $\textsl w_{[k]}$ and $\textsl v_{[k]}$:

\begin{align*} 
 & \frac{\mathrm{d}}{\mathrm{d}\textsl w_{[k]}}L_{2}  =   \\ 
&  \texttt{Enc}
 \begin{mbmatrix}
  s_{[1][1]}         &  \ldots  &  s_{[1][1]}     &   0   &  \ldots  & 0 
 \end{mbmatrix} 
 \hspace{3.5cm}
  \texttt{Enc}
 \begin{mbmatrix}
  s_{[1][c]}         &  \ldots  &  s_{[1][c]}     &   0   &  \ldots  & 0 
 \end{mbmatrix} 
 \\  
  & \hspace{1.74cm} \odot   \hspace{6.99cm}   \odot
 \\
&   \texttt{Enc}
 \begin{mbmatrix}
  v_{[1][1+k]}          &  \cdots  & v_{[1][1+k]}     &   0   &  \ldots  & 0 
 \end{mbmatrix} 
 \hspace{2.8cm}
  \texttt{Enc}
 \begin{mbmatrix}
  v_{[c][1+k]}          &  \cdots  & v_{[c][1+k]}     &   0   &  \ldots  & 0 
 \end{mbmatrix} 
 \\
 & \hspace{1.74cm} \odot   \hspace{3.74cm} + \ \ \cdots \ \  +   \hspace{1.74cm}   \odot  
 \\
&  \texttt{Enc}
 \begin{mbmatrix}
  \phi^{'}( z_{[1][k]} )        &  \cdots  & \phi^{'}( z_{[1][k]} )     &   0   &  \ldots  & 0 
 \end{mbmatrix} 
 \hspace{2.4cm}
  \texttt{Enc}
 \begin{mbmatrix}
  \phi^{'}( z_{[1][k]} )        &  \cdots  & \phi^{'}( z_{[1][k]} )     &   0   &  \ldots  & 0 
 \end{mbmatrix} 
 \\
 & \hspace{1.74cm} \odot  \hspace{6.99cm}    \odot
 \\
& \texttt{Enc}
 \begin{mbmatrix}
  x_{[1][0]}       &  \ldots  & x_{[1][d]}    &   0   &  \ldots  & 0 
 \end{mbmatrix} 
 \hspace{3.5cm}
  \texttt{Enc}
 \begin{mbmatrix}
  x_{[1][0]}       &  \ldots  & x_{[1][d]}    &   0   &  \ldots  & 0 
 \end{mbmatrix} \\
&   \hspace{6.49cm}    \\
&   \hspace{6.49cm} +   \\
&   \hspace{6.49cm}    \\
&  \texttt{Enc}
 \begin{mbmatrix}
  s_{[2][1]}         &  \ldots  &  s_{[2][1]}     &   0   &  \ldots  & 0 
 \end{mbmatrix} 
 \hspace{3.5cm}
  \texttt{Enc}
 \begin{mbmatrix}
  s_{[2][c]}         &  \ldots  &  s_{[2][c]}     &   0   &  \ldots  & 0 
 \end{mbmatrix} 
 \\  
  & \hspace{1.74cm} \odot   \hspace{6.99cm}   \odot
 \\
&   \texttt{Enc}
 \begin{mbmatrix}
  v_{[1][1+k]}          &  \cdots  & v_{[1][1+k]}     &   0   &  \ldots  & 0 
 \end{mbmatrix} 
 \hspace{2.8cm}
  \texttt{Enc}
 \begin{mbmatrix}
   v_{[c][1+k]}          &  \cdots  & v_{[c][1+k]}     &   0   &  \ldots  & 0 
 \end{mbmatrix} 
 \\
 & \hspace{1.74cm} \odot   \hspace{3.74cm} + \ \ \cdots \ \  +   \hspace{1.74cm}   \odot  
 \\
&  \texttt{Enc}
 \begin{mbmatrix}
  \phi^{'}( z_{[2][k]} )        &  \cdots  & \phi^{'}( z_{[2][k]} )     &   0   &  \ldots  & 0 
 \end{mbmatrix} 
 \hspace{2.4cm}
  \texttt{Enc}
 \begin{mbmatrix}
  \phi^{'}( z_{[2][k]} )        &  \cdots  & \phi^{'}( z_{[2][k]} )     &   0   &  \ldots  & 0 
 \end{mbmatrix} 
 \\
 & \hspace{1.74cm} \odot  \hspace{6.99cm}    \odot
 \\
& \texttt{Enc}
 \begin{mbmatrix}
  x_{[1][0]}       &  \ldots  & x_{[1][d]}    &   0   &  \ldots  & 0 
 \end{mbmatrix} 
 \hspace{3.5cm}
  \texttt{Enc}
 \begin{mbmatrix}
  x_{[1][0]}       &  \ldots  & x_{[1][d]}    &   0   &  \ldots  & 0 
 \end{mbmatrix} \\
 &   \hspace{6.49cm} +   \\
 &   \hspace{6.57cm} \vdots   \\
 &   \hspace{6.49cm} +   \\
&  \texttt{Enc}
 \begin{mbmatrix}
  s_{[n][1]}         &  \ldots  &  s_{[n][1]}     &   0   &  \ldots  & 0 
 \end{mbmatrix} 
 \hspace{3.5cm}
  \texttt{Enc}
 \begin{mbmatrix}
  s_{[n][c]}         &  \ldots  &  s_{[n][c]}     &   0   &  \ldots  & 0 
 \end{mbmatrix} 
 \\  
  & \hspace{1.74cm} \odot   \hspace{6.99cm}   \odot
 \\
&   \texttt{Enc}
 \begin{mbmatrix}
  v_{[1][1+k]}          &  \cdots  & v_{[1][1+k]}     &   0   &  \ldots  & 0 
 \end{mbmatrix} 
 \hspace{2.8cm}
  \texttt{Enc}
 \begin{mbmatrix}
 v_{[c][1+k]}          &  \cdots  & v_{[c][1+k]}     &   0   &  \ldots  & 0 
 \end{mbmatrix} 
 \\
 & \hspace{1.74cm} \odot   \hspace{3.74cm} + \ \ \cdots \ \  +   \hspace{1.74cm}   \odot  
 \\
&  \texttt{Enc}
 \begin{mbmatrix}
  \phi^{'}( z_{[n][k]} )        &  \cdots  & \phi^{'}( z_{[n][k]} )     &   0   &  \ldots  & 0 
 \end{mbmatrix} 
 \hspace{2.4cm}
  \texttt{Enc}
 \begin{mbmatrix}
  \phi^{'}( z_{[n][k]} )        &  \cdots  & \phi^{'}( z_{[n][k]} )     &   0   &  \ldots  & 0 
 \end{mbmatrix} 
 \\
 & \hspace{1.74cm} \odot  \hspace{6.99cm}    \odot
 \\
& \texttt{Enc}
 \begin{mbmatrix}
  x_{[1][0]}       &  \ldots  & x_{[1][d]}    &   0   &  \ldots  & 0 
 \end{mbmatrix} 
 \hspace{3.5cm}
  \texttt{Enc}
 \begin{mbmatrix}
  x_{[1][0]}       &  \ldots  & x_{[1][d]}    &   0   &  \ldots  & 0 
 \end{mbmatrix} , \\
\end{align*} 
 
  and

\begin{align*} 
 & \frac{\mathrm{d}}{\mathrm{d}\textsl v_{[k]}}L_{2}  = \\
 &  \texttt{Enc}
 \begin{mbmatrix}
  s_{[1][k]}     &   s_{[1][k]}     &  \ldots  &  s_{[1][k]}     &   0   &  \ldots  & 0 
 \end{mbmatrix}  
 \odot
   \texttt{Enc}
 \begin{mbmatrix}
  1      &      \phi( z_{[1][1]} )   &   \phi( z_{[1][2]} )   &  \cdots  & \phi( z_{[1][m]} )     &   0   &  \ldots  & 0 
 \end{mbmatrix}    \\
 &   \hspace{5.74cm} +  \\
 &  \texttt{Enc}
 \begin{mbmatrix}
  s_{[2][k]}    &   s_{[2][k]}    &  \ldots  & s_{[2][k]}    &   0   &  \ldots  & 0
 \end{mbmatrix} 
 \odot
   \texttt{Enc}
 \begin{mbmatrix}
  1      &      \phi( z_{[2][1]} )    &   \phi( z_{[2][2]} )   &  \cdots  & \phi( z_{[2][m]} )    &   0   &  \ldots  & 0
 \end{mbmatrix}     \\
&   \hspace{5.74cm} +             \\
&   \hspace{5.83cm} \vdots        \\
&   \hspace{5.74cm} +             \\
&  \texttt{Enc}
 \begin{mbmatrix}
  s_{[n][k]}     &  s_{[n][k]}    &  \ldots  & s_{[n][k]}   &   0   &  \ldots  & 0 
 \end{mbmatrix} 
 \odot
   \texttt{Enc}
 \begin{mbmatrix}
 1      &      \phi( z_{[n][1]} )    &   \phi( z_{[n][2]} )   &  \cdots  & \phi( z_{[n][m]} )    &   0   &  \ldots  & 0 
 \end{mbmatrix} .
\end{align*} 

Since the $\texttt{Double Volley Revolver}$ method only requires one of the two matrices to be transposed before encryption, and the expressions $\frac{\mathrm{d}}{\mathrm{d}\textsl w_{[k]}}L_{2}$ and $\frac{\mathrm{d}}{\mathrm{d}\textsl v_{[k]}}L_{2}$ happen to meet this requirement during matrix multiplication, we are able to carry out the homomorphic evaluation of the whole pipeline for homomorphic 3-layer NN training:   $\textsl w_{[k]} = \textsl w_{[k]} - \eta \cdot \frac{\mathrm{d}}{\mathrm{d}\textsl w_{[k]}}L_{2}$ and $\textsl v_{[k]} = \textsl v_{[k]} - \eta \cdot  \frac{\mathrm{d}}{\mathrm{d}\textsl v_{[k]}}L_{2}$, where the learning rate $\eta$ is set to $0.01$.

$L_2$ regularization, also referred to as Ridge Regression, introduces a penalty term proportional to the square of the model's parameters. This approach encourages the utilization of all parameters while minimizing their magnitudes, resulting in a less complex model that is less susceptible to overfitting. It is straightforward that $L_2$ regularization can be used in this encoding method by initializing the $\frac{\mathrm{d}}{\mathrm{d}\textsl w_{[k]}}L_{2}$ and $\frac{\mathrm{d}}{\mathrm{d}\textsl v_{[k]}}L_{2}$ with their corresponding L2 gradient components:

\begin{align*}
& \frac{\mathrm{d}}{\mathrm{d}\textsl w_{[k]}}L_{2} =
& \lambda \cdot \texttt{Enc}
\begin{bmatrix}
w_{[k][0]} & w_{[k][1]} & \ldots & w_{[k][d]} & 0 & \ldots & 0
\end{bmatrix}
+ \cdots ,
\end{align*}

and

\begin{align*}
& \frac{\mathrm{d}}{\mathrm{d}\textsl v_{[k]}}L_{2} =
& \lambda \cdot \texttt{Enc}
\begin{bmatrix}
v_{[k][0]} & v_{[k][1]} & \ldots & v_{[k][m]} & 0 & \ldots & 0
\end{bmatrix}
+ \cdots ,
\end{align*}
where $\lambda$ represents the $L_2$ regularization parameter.

\subsection{ For Regression Problems } 
The mean squared error (MSE) loss function can be used in conjunction with this encoding method for regression tasks. Interestingly, the special case of the $L_2$ loss function for a dataset containing only a single class label has the same formulation as the MSE loss function.

\subsection{ Approximated Activation Functions }
The first variant of the SLE loss function requires approximated activation functions to replace the sigmoid function with polynomials. Various methods~\cite{chiang2022polynomial, kim2018secure} can be employed to accomplish this task. For example, both Python and MATLAB (Octave) offer a function called $\texttt{polyfit}$, which can be used to approximate functions with polynomials using the least-square approach.

In our implementation, the activation function in the hidden layer is simply set to the square function.

\section{Experiments}
The C++ source code required to carry out the experiments described in this section is openly accessible at: \href{https://github.com/petitioner/HE.NNtraining}{$\texttt{https://github.com/petitioner/HE.NNtraining}$} .

\paragraph{Implementation}

We implemented vanilla gradient descent using the $L_2$ loss function, solely relying on homomorphic encryption (HE), with the $\texttt{HEAAN}$ library. All the experiments involving ciphertexts were performed on a public cloud with $64$ virtual CPUs and $192$ GB of RAM.

We use the entire set of $150$ Iris examples for training and employ a fixed learning rate of $0.01$. The complete Iris dataset $X$ and its corresponding one-hot encoded labels $Y$ are encrypted into two distinct ciphertexts. Moreover,  for each row of the weight matrices $W$ and $V$, we encrypt its multiple repetitions  into a single ciphertext:

\begin{align*}
  X & \xmapsto{ \text{ encoding \& encryption } } 
   Enc 
 \begin{bmatrix}
 x_{[1][0]}    &   x_{[1][1]}   &  \cdots  & x_{[1][d]}   \\
 x_{[2][0]}    &   x_{[2][1]}   &  \cdots  & x_{[2][d]}   \\
 \vdots    &   \vdots   &  \ddots  & \vdots   \\
 x_{[n][0]}    &   x_{[n][1]}   &  \cdots  & x_{[n][d]}   \\
 \end{bmatrix},   \\     
\textsl Y & \xmapsto{ \text{ encoding \& encryption } }
Enc
 \begin{bmatrix}
 y_{[1][1]}    &   y_{[1][2]}   &  \cdots  & y_{[1][c]}   \\
 y_{[2][1]}    &   y_{[2][2]}   &  \cdots  & y_{[2][c]}   \\
 \vdots    &   \vdots   &  \ddots  & \vdots   \\
 y_{[n][1]}    &   y_{[n][2]}   &  \cdots  & y_{[n][c]}   \\
 \end{bmatrix}, \\
 W & \xmapsto{ \text{ encoding \& encryption } }   
 Enc \begin{bmatrix}
 w_{[1][0]}    &   w_{[1][1]}   &  \cdots  & w_{[1][d]}   \\
 w_{[1][0]}    &   w_{[1][1]}   &  \cdots  & w_{[1][d]}   \\
 \vdots    &   \vdots   &  \ddots  & \vdots   \\
 w_{[1][0]}    &   w_{[1][1]}   &  \cdots  & w_{[1][d]}   \\
 \end{bmatrix} , \cdots , 
 Enc \begin{bmatrix}
 w_{[m][0]}    &   w_{[m][1]}   &  \cdots  & w_{[m][d]}   \\
 w_{[m][0]}    &   w_{[m][1]}   &  \cdots  & w_{[m][d]}   \\
 \vdots    &   \vdots   &  \ddots  & \vdots   \\
 w_{[m][0]}    &   w_{[m][1]}   &  \cdots  & w_{[m][d]}   \\
 \end{bmatrix} , \\
 V & \xmapsto{ \text{ encoding \& encryption } }     
 Enc \begin{bmatrix}
 v_{[1][0]}    &   v_{[1][1]}   &  \cdots  & v_{[1][m]}   \\
 v_{[1][0]}    &   v_{[1][1]}   &  \cdots  & v_{[1][m]}   \\
 \vdots    &   \vdots   &  \ddots  & \vdots   \\
 v_{[1][0]}    &   v_{[1][1]}   &  \cdots  & v_{[1][m]}   \\
 \end{bmatrix} , \cdots ,
 Enc \begin{bmatrix}
 v_{[c][0]}    &   v_{[c][1]}   &  \cdots  & v_{[c][m]}   \\
 v_{[c][0]}    &   v_{[c][1]}   &  \cdots  & v_{[c][m]}   \\
 \vdots    &   \vdots   &  \ddots  & \vdots   \\
 v_{[c][0]}    &   v_{[c][1]}   &  \cdots  & v_{[c][m]}   \\
 \end{bmatrix}.
\end{align*} 

We selected the following parameters for the $\texttt{HEAAN}$ library: $logN = 16$, $logQ = 990$, $logp = 30$, $slots = 32768$, which collectively ensure a security level of $\lambda = 128$. For detailed information about these parameters, please refer to \cite{IDASH2018Andrey}. We did not employ bootstrapping to refresh the weight ciphertexts, which limits our algorithm to only perform $2$ iterations. Each iteration takes approximately $52$ minutes to complete. The maximum runtime memory usage under these conditions is around $60$ GB. The dataset of 150 Iris examples is encrypted into a single ciphertext. In addition, the one-hot encoded labels $\textsl Y$ are encrypted into a single ciphertext. The weight matrix $W$ is encrypted into $120$ ciphertexts, with each ciphertext encrypting multiple repetitions of each row. We initialized the weight matrices $W$ and $V$ with a normal distribution having a mean of $0.0$ and a standard deviation of $0.05$.


\section{Conclusion}

In this work, we implemented privacy-preserving 3-layer NN training using solely homomorphic encryption (HE) techniques. However, the current low-level encoding method within the framework of Double Volley Revolver is not well-suited for incorporating bootstrapping. Further research is necessary to investigate the integration of bootstrapping into our approach using an alternative low-level implementation.

\bibliography{HE.NNtraining}
\bibliographystyle{apalike}  

\end{document}